\documentclass[aps,rpl,preprint,groupedaddress, showpacs]{revtex4}
\usepackage{epsfig, amsmath, amssymb}

\newcommand{\bra}[1]{\langle #1|}
\newcommand{\ket}[1]{|#1\rangle}
\newcommand{\braket}[2]{\langle #1|#2\rangle}
\newcommand{\ketbra}[2]{|#1\rangle\langle #2|}

\begin{document}

\title{Anisotropy and Magnetic Field Effects on the Genuine Multipartite Entanglement of Multi-Qubit Heisenberg {\it XY} Chains}
\author{Chang Chi Kwong and Ye Yeo}
\affiliation{Department of Physics, National University of Singapore, 10 Kent Ridge Crescent, Singapore 119260, Singapore}

\begin{abstract}
It has been shown that, for the two-qubit Heisenberg $XY$ model, anisotropy and magnetic field may together be used to produce entanglement for any finite temperature by adjusting the external magnetic field beyond some finite critical strength.  This interesting result arises from an analysis employing the Wootters concurrence, a computable measure of entanglement for two-qubit states.  Recently, Mintert {\em et al.} proposed generalizations of Wootters concurrence for multipartite states.  These MKB concurrences possess a mathematical property that enables one to understand the origin of this characteristic behavior.  Here, we first study the effect of anisotropy and magnetic field on the multipartite thermal entanglement of a four-qubit Heisenberg $XY$ chain using the MKB concurrences.  We show that this model exhibits characteristic behavior similar to that of the two-qubit model.  In addition, we show that this can again be understood using the same mathematical property.  Next, we show that the six-qubit Heisenberg $XY$ chain possesses properties necessary for it to have the characteristic behavior too.  Most importantly, it is possible to directly measure the multipartite MKB concurrences of pure states.  This may provide an experimental verification of our conjecture that for a Heisenberg $XY$ chain of any even number of qubits, it is always possible to obtain non-zero genuine multipartite entanglement at any finite temperature by applying a sufficiently large magnetic field.
\end{abstract}

\maketitle

\section{Introduction}
The one-dimensional Heisenberg models have been extensively studied in solid state physics (see references in \cite{arnesen01}).  Interest in these models has been revived lately by several proposals for realizing quantum computation \cite{Loss} and information processing \cite{Imamoglu} using quantum dots (localized electron spins) as qubits.  An intriguing phenomenon in such quantum systems with more than one component is entanglement.  It refers to the non-classical correlations that exist among these components.  Due to this nature, quantum entanglement has recently been recognized as an indispensable physical resource for performing classically impossible information processing tasks; such as quantum computation, teleportation \cite{bennett93}, key distribution \cite{ekert91} and superdense coding \cite{bennett92}.  In any physically realistic consideration of quantum-information-processing device, it is important to take into account the fact that it would be in thermal contact with some heat bath.  Consequently, entanglement in interacting Heisenberg spin systems at finite temperatures has been investigated by a number of authors (see, e.g., Ref.\cite{Osborne} and references therein).  The state of a typical solid state system at thermal equilibrium (temperature $T$) is $\chi = {\rm e}^{-\beta H}/Z$, where $H$ is the Hamiltonian, $Z = {\rm tr}\,{\rm e}^{-\beta H}$ is the partition function, and $\beta = 1/kT$, where $k$ is the Boltzmann's constant.  The entanglement associated with the thermal state $\chi$ is referred to as the thermal entanglement \cite{arnesen01}.

To achieve a quantitative understanding of the role of entanglement in the field of quantum information science, it is necessary to quantify the amount of entanglement that is associated with a given state.  Several entanglement measures have been proposed.  One famous example is the Wootters concurrence \cite{wootters98}.  Consider a two-qubit state $\rho$, the Wootters concurrence
\begin{equation}
{\cal C}_W[\rho] \equiv \max\{\lambda_1 - \lambda_2 - \lambda_3 - \lambda_4,\ 0\},
\end{equation}
where $\lambda_k$ $(k = 1, 2, 3, 4)$ are the square roots of the eigenvalues in decreasing order of magnitude of the spin-flipped density-matrix operator $R = \rho(\sigma^y \otimes \sigma^y)\rho^*(\sigma^y \otimes \sigma^y)$, the asterisk indicates complex conjugation.  Wootters concurrence is closely related to the entanglement of formation \cite{bennett96}.  It appears as an auxillary function that has to be evaluated when computing the entanglement of formation.  Recently, Mintert, K\'{u}s and Buchleitner (MKB) \cite{mintert05, dd06} proposed generalizations of Wootters concurrence for multipartite quantum systems, which can be evaluated efficiently for arbitrary mixed states.  This is an extremely important development in our investigation of multipartite entanglement, a subject we have yet to achieve a complete understanding.

With the availability of such a good and computable measure of entanglement for systems of two qubits, many thoroughly analyzed the thermal entanglement in two-qubit Heisenberg models in terms of thermal concurrence \cite{arnesen01, Wang1, Wang2, Gunlycke, kamta02, Anteneodo, Zhou}.  The recent breakthroughs in the experimental physics of double quantum dot (see, e.g., Ref.\cite{Chen}) shows that these studies are worthwhile pursuits.  Interestingly, Kamta {\em et al.} \cite{kamta02} showed that the anisotropy and the magnetic field may together be used to control the extent of thermal concurrence in a two-qubit Heisenberg $XY$ model and, especially, to produce entanglement for any finitely large $T$, by adjusting the external magnetic field beyond some finitely large critical strength.  Such robustness is absent in the case of two-qubit Heisenberg $XX$ model.  A natural question is whether thermal entanglement of a Heisenberg $XY$ chain with more qubits has similar behavior.  This question is not only of practical importance but also of fundamental importance (see, for instance, Ref.\cite{wang06}).  In this paper, we provide answers for Heisenberg $XY$ chains with any even number of qubits by analyzing the MKB concurrences associated with their thermal states.

Our paper is organized as follows.  Since the results of MKB \cite{mintert05, dd06} is crucial to our analysis, we will present them in the following section.  However, we do so explicitly in the context of direct relevance to our paper, i.e., $n$-qubits.  Specifically, when $n = 2$, the MKB concurrence coincides with the Wootters concurrence.  It thus allows us to understand the special property of the thermal entanglement of the two-qubit Heisenberg $XY$ model, in the light of a mathematical property of the MKB concurrences (see Eq.(\ref{approx})).  This together with a discussion of the two-qubit Heisenberg $XY$ model \cite{kamta02} will be given in Section III.  We show, in Section IV, that the thermal entanglement of the four-qubit Heisenberg $XY$ chain has behavior similar to the two-qubit case.  Namely, anisotropy and magnetic field may together be used to produce {\em genuine} multipartite entanglement \cite{osterloh05, Osterloh05} for any finitely large $T$ by adjusting the external magnetic field beyond some finitely large critical strength.  This can again be understood in the light of the same mathematical property of MKB concurrences.  Our results agree with those in Ref.\cite{rossignoli05}, which employed bipartite entanglement measure.  In general, it is not sufficient to only consider bipartite entanglement.  For instance, the Greenberger-Horne-Zeilinger (GHZ) state \cite{Greenberger} is a state with genuine multipartite entanglement but yields zero entanglement between one particle and any other particle.  On the other hand, the W state \cite{Zeilinger} is one where every particle is entangled with every other particle, but it has no genuine multipartite entanglement \cite{mintert05, osterloh05}.  Multipartite entanglement surely is more interesting.  Therefore, our study complements theirs.  However, we must emphasize that our analysis more importantly demonstrates that the thermal entanglement of the Heisenberg $XY$ chains have the characteristic behavior due to two reasons.  First, in the presence of a finitely large external magnetic field, the ground state of the model has non-zero genuine multipartite entanglement.  Second, for non-zero temperatures, this ground state is the highest weight state (with weight $\approx 1$, i.e., the thermal state is almost a {\em pure} state) when a finitely large enough magnetic field is applied.  In Section V, we show that this is indeed the case for the six-qubit Heisenberg $XY$ chain.  We thus have firm mathematical basis (Eq.(\ref{approx})) to establish that the thermal entanglement in this case will exhibit similar characteristic behavior.  Most important of all, the MKB concurrences may be directly measured for multipartite pure states \cite{Walborn, aolita06}.  It is thus possible to experimentally verify our conjecture that for any even $n$, it is always possible to obtain non-zero genuine multipartite entanglement at any finite temperature by applying a sufficiently large magnetic field.  Further discussions of this possibility and a summary of our results will be presented in the concluding Section VI.


\section{The MKB Concurrences}
Consider an $n$-qubit pure state $\ket{\psi} \in {\cal H}_1 \otimes {\cal H}_2 \otimes \cdots \otimes {\cal H}_n$, the MKB concurrence is defined as the expectation value of a Hermitean operator $A$ that acts on two copies of the state \cite{mintert05}:
\begin{equation}
{\cal C}[\ket{\psi}\bra{\psi}] \equiv \sqrt{\bra{\psi}\otimes\bra{\psi}A\ket{\psi}\otimes\ket{\psi}}.
\end{equation}
In general, $A$ could have the form
\begin{equation}\label{defA}
A \equiv \sum_{\{s_{j_i}=\pm\}^+} p_{s_{1_i}s_{2_i} \cdots s_{n_i}}
P^{(1)}_{s_{1_i}} \otimes P^{(2)}_{s_{2_i}} \otimes \cdots \otimes P^{(n)}_{s_{n_i}},
\end{equation}
where $p_{s_{1_i}s_{2_i} \cdots s_{n_i}} \geq 0$,
\begin{eqnarray}
P^{(j)}_+ & \equiv & \frac{1}{2}(\Pi^+_0 + \Pi^+_1 + \Pi^-_0), \nonumber \\
P^{(j)}_- & \equiv & \frac{1}{2}\Pi^-_1
\end{eqnarray}
are projectors onto the symmetric and antisymmetric subspaces of $\mathcal{H}_j\otimes\mathcal{H}_j$.  Here, $\Pi^{\pm}_0 \equiv (\ket{00} \pm \ket{11})(\bra{00} \pm \bra{11})$, $\Pi^{\pm}_1 \equiv (\ket{01} \pm \ket{10})(\bra{01} \pm \bra{10})$ and $\{\ket{0}, \ket{1}\}$ is an orthonormal basis of $\mathcal{H}_j$.  The summation in Eq.(\ref{defA}) is performed over the set $\{s_{j_i}=\pm\}^+$, which contains all $n$-long strings of $+$'s and $-$'s with even number of $-$'s.  The superscript $+$ indicates that the string with $n$ $+$'s is not included in the sum.  This is because for separable states, its expectation value in the symmetric twofold copy is non-zero. Terms with odd number of $P^{(j)}_-$'s are naturally excluded in the sum since their expectation values in the twofold copy states is always zero.  

By choosing the value of all the $p_{s_{1_i}s_{2_i} \cdots s_{n_i}}$'s in Eq.(\ref{defA}) to be 4, an entanglement monotone ${\cal C}_n$ can be obtained.  The resulting operator $A_n$ can equivalently be written as $4(I - P^{(1)}_+ \otimes P^{(2)}_+ \otimes \cdots \otimes P^{(n)}_+)$ \cite{aolita06}.  The concurrence ${\cal C}_n$  of an $n$-qubit pure state $|\psi\rangle$ can then be written as
\begin{equation}
{\cal C}_n[\ket{\psi}\bra{\psi}] = 2^{1-n/2}\sqrt{(2^n-2)\braket{\psi}{\psi}-\sum_i\mathrm{Tr}\rho_i^2}.
\end{equation}
The above summation runs over all $(2^n-2)$ reduced density operators $\rho_i$ of the state $\ket{\psi}$.  ${\cal C}_n$ accounts for all possible types of entanglement in a state and takes the value zero if and only if the state is fully separable.

For even number $n$ of qubits, it is possible to define an MKB concurrence ${\cal C}^{(n)}$ that detects multipartite entanglement, by choosing the operator $A = A^{(n)} \equiv 2^nP^{(1)}_- \otimes P^{(2)}_- \otimes \cdots \otimes P^{(n)}_-$.  We note that when $n = 2$,
\begin{equation}\label{MKBWequiv}
{\cal C}^{(2)}[\ket{\psi}\bra{\psi}] = |\langle\psi^*|\sigma^y \otimes \sigma^y|\psi\rangle|  = {\cal C}_W[\ket{\psi}\bra{\psi}].
\end{equation}
That is, the MKB concurrence ${\cal C}^{(2)}$ coincides with the Wootters concurrence.  And, for $n = 4$, we have
\begin{equation}
{\cal C}^{(4)}[\ket{\psi}\bra{\psi}] = |\bra{\psi^*}\sigma^y \otimes \sigma^y \otimes \sigma^y \otimes \sigma^y\ket{\psi}|,
\end{equation}
which is also an entanglement monotone \cite{dd06}.  There is obviously no equivalent definition for the case of odd number of qubits since the expectation value of $A^{(n)}$ are always zero for odd $n$.

The MKB concurrence for a mixed state $\rho$ of $n$ qubits can be obtained via the convex roof construction:
\begin{equation}\label{convexroof}
{\cal C}[\rho] \equiv \inf\left\{\sum_ip_i{\cal C}[\ket{\psi_i}\bra{\psi_i}],\ \rho = \sum_ip_i|\psi_i\rangle\langle\psi_i|\right\},
\end{equation}
where the infimum is taken over all possible pure state decompositions of $\rho$.  To evaluate ${\cal C}[\rho]$, consider the spectral decompositions $\rho = \sum_i\ketbra{\tilde{\phi}_i}{\tilde{\phi}_i}$ and $A = \sum_\alpha\ketbra{\tilde{\chi}_\alpha}{\tilde{\chi}_\alpha}$, where the eigenstates $\ket{\tilde{\phi}_i}$ and $\ket{\tilde{\chi}_\alpha}$ are subnormalized such that their norms squared are the eigenvalues corresponding to the states.  If $r$ is the rank of the operator $A$, it is possible to define $r$ complex symmetric matrices $T^\alpha$ with elements $T^\alpha_{jk} \equiv \bra{\tilde{\phi}_j} \otimes \braket{\tilde{\phi}_k}{\tilde{\chi}^\alpha}$.  And, Eq.(\ref{convexroof}) becomes
\begin{equation}
{\cal C}[\rho] = \inf_V\sum_i\sqrt{\sum_\alpha|[VT^\alpha V^T]_{ii}|^2}
\end{equation} 
where the infimum is now taken over the set of left unitary matrices $V$.

It can be shown that the following inequality holds \cite{mintert05}:
\begin{equation}\label{ineqC}
{\cal C}[\rho] \geq \inf_V\sum_i|[V\tau V^T]_{ii}|.
\end{equation}
Here, the matrix $\tau$ is defined to be $\sum_\alpha z_\alpha T^\alpha$ in terms of arbitrary complex numbers $z_\alpha$ satisfying only the condition that $\sum_\alpha|z_\alpha|^2 = 1$.  An algebraic solution of the inequality Eq.(\ref{ineqC}) is given in Ref.\cite{mintert05} to be $\max\{0,\ \lambda_1 - \sum_{j>1}\lambda_j\}$ where $\lambda_i$'s are singular values of $\tau$ written in decreasing order.  For ${\cal C}^{(n)}$, when $A = A^{(n)}$ is of rank 1, $T^1 = \tau$ and the lower bound in Eq.(\ref{ineqC}) turns out to be the exact value of ${\cal C}^{(n)}$.

In general, an optimization over $z_\alpha$ is also necessary to obtain the optimal lower bound for ${\cal C}[\rho]$.  However, it is possible to obtain a good approximation to ${\cal C}[\rho]$ by approximating $\tau$ with a matrix whose elements \cite{mintert05}
\begin{equation}\label{approx}
\tau_{ij} \approx 
\frac{\bra{\tilde{\phi}_1}\otimes\bra{\tilde{\phi}_1}A\ket{\tilde{\phi}_i}\otimes\ket{\tilde{\phi}_j}}
{\sqrt{\bra{\tilde{\phi}_1}\otimes\bra{\tilde{\phi}_1}A\ket{\tilde{\phi}_1}\otimes\ket{\tilde{\phi}_1}}},
\end{equation}
where $\ket{\tilde{\phi}_1}$ is the eigenstate of $\rho$ with the largest eigenvalue.  This is the mathematical property of MKB concurrences that will play a critical role in our understanding of the characteristic behavior of the thermal entanglement of Heisenberg $XY$ models.  We will first illustrate this explicitly with the two-qubit Heisenberg $XY$ chain in the next section.


\section{Two-qubit Heisenberg $XY$ Model}
The Hamiltonian $H_2$ for the anisotropic two-qubit Heisenberg $XY$ model in an external magnetic field $B_m \equiv \eta J$ ($\eta$ is a real number) along the $z$ axis is
\begin{equation}
H_2 = \frac{1}{2}(1 + \gamma)J\sigma^x_1 \otimes \sigma^x_2 + \frac{1}{2}(1 - \gamma)J\sigma^y_1 \otimes \sigma^y_2 + \frac{1}{2}B_m(\sigma^z_1 \otimes I_2 + I_1 \otimes \sigma^z_2),
\end{equation}
where $I_j$ is the identity matrix and $\sigma^x_j$, $\sigma^y_j$, $\sigma^z_j$ are the Pauli matrices at site $j = 1, 2$.  The parameter $-1 \leq \gamma \leq 1$ measures the anisotropy of the system and equals $0$ for the isotropic $XX$ model \cite{Wang1} and $\pm 1$ for the Ising model \cite{Gunlycke}.  $(1 + \gamma)J$ and $(1 - \gamma)J$ are real coupling constants for the spin interaction.  The model is said to be antiferromagnetic for $J > 0$ and ferromagnetic for $J < 0$.  The thermal concurrence associated with the thermal state $\chi_2$, Eq.(\ref{thermalstate2}), can be derived from Eq.(\ref{Wcon}).  It is invariant under the substitutions $\eta \longrightarrow -\eta$, $\gamma \longrightarrow -\gamma$, and $J \longrightarrow -J$.  Therefore, we restrict our considerations to $\eta \geq 0$, $0 \leq \gamma \leq 1$, and $J > 0$.

The eigenvalues and eigenkets of $H_2$ are given by \cite{kamta02}
\begin{eqnarray}
H_2|\Phi^0\rangle & = & {\cal B} |\Phi^0\rangle, \nonumber \\
H_2|\Phi^1\rangle & = & J        |\Phi^1\rangle,\ \nonumber \\
H_2|\Phi^2\rangle & = & -J       |\Phi^2\rangle,\ \nonumber \\
H_2|\Phi^3\rangle & = & -{\cal B}|\Phi^3\rangle,
\end{eqnarray}
where ${\cal B} \equiv \sqrt{B^2_m + \gamma^2J^2} = \sqrt{\eta^2 + \gamma^2}J$,
\begin{eqnarray}
|\Phi^0\rangle & = & \frac{1}{\sqrt{({\cal B} + B_m)^2 + \gamma^2J^2}}
[({\cal B} + B_m)|00\rangle + \gamma J|11\rangle], \nonumber \\
|\Phi^1\rangle & = & \frac{1}{\sqrt{2}}[|01\rangle + |10\rangle], \nonumber \\
|\Phi^2\rangle & = & \frac{1}{\sqrt{2}}[|01\rangle - |10\rangle], \nonumber \\
|\Phi^3\rangle & = & \frac{1}{\sqrt{({\cal B} - B_m)^2 + \gamma^2J^2}}
[({\cal B} - B_m)|00\rangle - \gamma J|11\rangle].
\end{eqnarray}
The Wootters concurrence associated with the eigenkets, $|\Phi^0\rangle$ and $|\Phi^3\rangle$, are given by $\frac{\gamma}{\sqrt{\eta^2 + \gamma^2}}$.  Hence, they represent entangled states when $\gamma \not= 0$.  We note that when $\eta = 0$, $|\Phi^0\rangle$ and $|\Phi^3\rangle$ reduce to $(|00\rangle + |11\rangle)/\sqrt{2}$ and $(|00\rangle - |11\rangle)/\sqrt{2}$ respectively, so that the eigenstates are the four maximally entangled Bell states: $|\Psi^0_{\rm Bell}\rangle$, $|\Psi^1_{\rm Bell}\rangle$, $|\Psi^2_{\rm Bell}\rangle$, and $|\Psi^3_{\rm Bell}\rangle$.  And, in the limit of large $\eta$,
\begin{equation}\label{largeeta}
{\cal C}_W[|\Phi^0\rangle\langle\Phi^0|] = {\cal C}_W[|\Phi^3\rangle\langle\Phi^3|] \approx \gamma\eta^{-1},
\end{equation}
only going to zero asympotically when $\eta$ is infinitely large.  In contrast, when $\gamma = 0$, $|\Phi^0\rangle = |00\rangle$ and $|\Phi^3\rangle = |11\rangle$ are product states with eigenvalues $\eta J$ and $-\eta J$ respectively, though $|\Phi^1\rangle$ and $|\Phi^2\rangle$ remain the same \cite{Wang1}.

For the above system in thermal equilibrium at temperature $T$, its state is described by the density operator
\begin{eqnarray}\label{thermalstate2}
\chi_2 & = & \sum^3_{i = 0}w_i|\Phi^i\rangle\langle\Phi^i| \nonumber \\
       & = & \frac{1}{Z_2}[e^{-\beta{\cal B}}|\Phi^0\rangle\langle\Phi^0| + e^{-\beta J}|\Phi^1\rangle\langle\Phi^1| 
                         + e^{\beta J}|\Phi^2\rangle\langle\Phi^2| + e^{\beta{\cal B}}|\Phi^3\rangle\langle\Phi^3|],
\end{eqnarray}
where the partition function $Z_2 = 2\cosh\beta{\cal B} + 2\cosh\beta J$, the Boltzmann's constant $k \equiv 1$ from hereon, and $\beta = 1/T$.  After some straightforward algebra, we obtain
\begin{eqnarray}\label{Wcon}
\lambda_1 & = & \frac{1}{Z_2}e^{\beta J}, \nonumber \\
\lambda_2 & = & \frac{1}{Z_2}e^{-\beta J}, \nonumber \\
\lambda_3 & = & \frac{1}{Z_2}\sqrt{1 + \frac{2\gamma^2J^2}{{\cal B}^2}\sinh^2\beta{\cal B} + \frac{2\gamma J}{{\cal B}}\sqrt{1 + \frac{\gamma^2J^2}{{\cal B}^2}\sinh^2\beta{\cal B}}\sinh\beta{\cal B}}, \nonumber \\
\lambda_4 & = & \frac{1}{Z_2}\sqrt{1 + \frac{2\gamma^2J^2}{{\cal B}^2}\sinh^2\beta{\cal B} - \frac{2\gamma J}{{\cal B}}\sqrt{1 + \frac{\gamma^2J^2}{{\cal B}^2}\sinh^2\beta{\cal B}}\sinh\beta{\cal B}}.
\end{eqnarray}

In the zero-temperature limit, i.e., $\beta \longrightarrow \infty$, at which the system is in its ground state, Eq.(\ref{thermalstate2}) reduces to the following three possibilities.
\begin{description}
\item{(a)} $0 \leq \eta < \sqrt{1 - \gamma^2}$:
\begin{equation}
\chi_2 = \frac{1}{Z_2}[e^{\beta J}|\Phi^2\rangle\langle\Phi^2| + e^{\beta{\cal B}}|\Phi^3\rangle\langle\Phi^3|]
\longrightarrow |\Phi^2\rangle\langle\Phi^2|,
\end{equation}
with $Z_2 = e^{\beta J} + e^{\beta\cal B}$.  Equation (\ref{Wcon}) gives ${\cal C}_W[\chi_2] = 1$, its maximum value, in agreement with the fact that $|\Phi^2\rangle$ is a maximally entangled Bell state.
\item{(b)} $\eta = \sqrt{1 - \gamma^2}$:
\begin{equation}
\chi_2 \longrightarrow \frac{1}{2}[|\Phi^2\rangle\langle\Phi^2| + |\Phi^3\rangle\langle\Phi^3|].
\end{equation}
From Eq.(\ref{Wcon}), the above equally weighted mixture has
\begin{equation}
{\cal C}_W[\chi_2] = \frac{1}{2}(1 - \gamma).
\end{equation}
\item{(c)} $\eta > \sqrt{1 - \gamma^2}$:
\begin{equation}
\chi_2 \longrightarrow |\Phi^3\rangle\langle\Phi^3|,
\end{equation}
and Eq.(\ref{Wcon}) yields accordingly
\begin{equation}
{\cal C}_W[\chi_2] = \frac{\gamma}{\sqrt{\eta^2 + \gamma^2}}.
\end{equation}
\end{description}
Therefore, for a given $\gamma$, $\eta_{\rm critical} = \sqrt{1 - \gamma^2}$ marks the point of quantum phase transition (phase transition taking place at zero temperature due to variation of interaction terms in the Hamiltonian of a system \cite{arnesen01}).  For values of $\gamma$ other than $\gamma = \frac{1}{3}$, there is a sudden increase or decrease in ${\cal C}[\chi_2]$ at $\eta_{\rm critical}$, depending on whether $\gamma > \frac{1}{3}$ or $\gamma < \frac{1}{3}$, before decreasing to zero asymptotically, as $\eta$ is increased beyond the critical value $\eta_{\rm critical}$ \cite{kamta02}.  Here, we focus on the behavior of the model at non-zero temperatures and subject to magnetic field of appropriate strengths.

At non-zero temperatures, due to mixing, ${\cal C}_W[\chi_2]$ decreases to zero as the temperature $T$ is increased beyond some critical value.  In fact, as $T \rightarrow \infty$, the statistical weights $w_i \rightarrow 1/4$ for all $i$ and ${\cal C}_W[\chi_2] \rightarrow 0$. However, for a large but finite $T$,
\begin{equation}\label{highestweight}
w_3 = \frac{e^{\beta\cal B}}{Z_2} = \frac{1}{1 + e^{-2\beta\cal B} + e^{-\beta({\cal B} - J)} + e^{-\beta({\cal B} + J)}}
\end{equation}
can always be made as close to unity as possible by increasing the strength $\eta$ of the external magnetic field.  That is, when $\eta$ is large enough, only $\ket{\Phi^3}$ contributes significantly to the thermal state $\chi_2$.  The $\eta$ required for this to occur depends on $T$, larger $\eta$ for higher $T$.  This is always possible for finite $T$.

Next, we note that in the limit of large $\eta$,
\begin{equation}
\lambda_1 \approx \lambda_2 \approx 0 \approx \lambda_4,\ \lambda_3 \approx \gamma\eta^{-1}.
\end{equation} 
It follows that
\begin{equation}
{\cal C}_W[\chi_2] \approx \gamma\eta^{-1} \approx {\cal C}_W[|\Phi^3\rangle\langle\Phi^3|].
\end{equation}
Hence, for a finitely large $T$, the thermal concurrence of the system in a large enough magnetic field is very well approximated by the concurrence of $|\Phi^3\rangle$ in the same magnetic field.  In other words, the entanglement associated with the thermal state $\chi_2$ in this case is mainly due to that associated with the eigenstate $\ket{\Phi^3}$.

In summary, for any non-zero $T$ and an appropriate $\eta$, $|\Phi^3\rangle$ is the highest weight eigenstate.  It follows from Eq.(\ref{approx}) that this is the state which will mainly contribute to the MKB concurrence ${\cal C}^{(2)}[\chi_2]$ or equivalently the Wootters concurrence (see Eq.(\ref{MKBWequiv})).  An important point to note here is that $|\Phi^3\rangle$ has non-zero concurrence as long as $\eta$ is finite (see Eq.(\ref{largeeta})).  Another is that $\gamma \not= 0$, otherwise $|\Phi^3\rangle$ will be a product state with no entanglement.  These clearly explain the characteristic robustness of the thermal entanglement of the two-qubit Heisenberg $XY$ model described in Ref\cite{kamta02} - the system at finite $T$ can always be entangled provided large enough magnetic field is applied.  They also identify the necessary characteristic features for a model to exhibit such behavior.  We shall illustrate that this is indeed the case for the four-qubit Heisenberg $XY$ chain in the next section.


\section{Four-qubit Heisenberg $XY$ model}
The Hamiltonian $H_n$ for an anisotropic $n$-qubit Heisenberg $XY$ chain in an external magnetic field $B_m = \eta J$ along the $z$-axis is
\begin{equation}
H_n = \frac{J}{2}\sum^n_{j = 1}[(1 + \gamma)\sigma^x_j\sigma^x_{j + 1} + (1 - \gamma)\sigma^y_j\sigma^y_{j + 1} + \eta\sigma^z_j],
\end{equation} 
where the periodic boundary condition $\sigma^{\alpha}_{n+1} = \sigma^{\alpha}_1$ $(\alpha = x, y, z)$ applies.  Like in the two-qubit model, we consider the case when $\eta \geq 0$, $0 \leq \gamma \leq 1$, and $J > 0$.

In this section, we consider $n = 4$.  After some straightforward algebra, we obtain the eigenvalues and eigenvectors of $H_4$.  Firstly, we present
\begin{equation}
H_4|\Phi^{15}\rangle = -\omega^+J|\Phi^{15}\rangle,
\end{equation}
where
\begin{equation}
\omega^{\pm} \equiv \sqrt{2}\sqrt{[\eta^2 + 2(1 + \gamma^2)] \pm \sqrt{[\eta^2 + 2(1 + \gamma^2)]^2 - 8\eta^2}},
\end{equation}
and
\begin{eqnarray}
|\Phi^{15}\rangle 
& = & N^-_{\Omega}(\Omega^-_1|0000\rangle + \Omega^-_2|0011\rangle + \Omega^-_3|0101\rangle + \Omega^-_2|0110\rangle \nonumber \\
& & \,\,\,\,\,\,\, + \Omega^-_2|1001\rangle + \Omega^-_3|1010\rangle + \Omega^-_2|1100\rangle + |1111\rangle),
\end{eqnarray}
with $N^{\pm}_{\Omega} \equiv 1/\sqrt{1 + (\Omega^{\pm}_1)^2 + 4(\Omega^{\pm}_2)^2 + 2(\Omega^{\pm}_3)^2}$,
\begin{eqnarray}\label{Omega}
\Omega_1^\pm & = & \frac{(2\eta \pm \omega^+)(\omega^{+2}-8) - 8\gamma^2(\eta \pm \omega^+)}{8\gamma^2\eta}, \nonumber\\
\Omega_2^\pm & = & \frac{2\eta \pm \omega^+}{4\gamma}, \nonumber\\
\Omega_3^\pm & = & \pm\frac{2\eta \pm \omega^+}{\gamma\omega^+}.
\end{eqnarray}
$|\Phi^0\rangle$, which can be obtained from $|\Phi^{15}\rangle$ by substituting $\Omega^-_i$ $(i = 1, 2, 3)$ with $\Omega^+_i$ or $\Omega^-_i \rightarrow \Omega^+_i$, satisfies $H_4|\Phi^0\rangle = \omega^+J|\Phi^0\rangle$.  The MKB concurrences, as defined in Section II, for $\ket{\Phi^{15}}$ are given by
\begin{eqnarray}
{\cal C}^{(4)}[|\Phi^{15}\rangle\langle\Phi^{15}|] & = & 
2\frac{\Omega^-_1+2(\Omega_2^-)^2+(\Omega^-_3)^2}{1+(\Omega_1^-)^2+4(\Omega_2^-)^2+2(\Omega_3^-)^2}, \nonumber \\
{\cal C}_4[|\Phi^{15}\rangle\langle\Phi^{15}|] & = & \frac{\sqrt{
7[2(\Omega^-_2)^2 + (\Omega^-_3)^2]^2 + 2[4(\Omega^-_1)^2 - \Omega^-_1 + 4][2(\Omega^-_2)^2 + (\Omega^-_3)^2] + 7(\Omega^-_1)^2}}
{1 + (\Omega^-_1)^2 + 4(\Omega^-_2)^2 + 2(\Omega^-_3)^2}. \nonumber \\
\end{eqnarray}
In the limit of large $\eta$,
\begin{eqnarray}
\omega^+   & \approx & 2\eta + \frac{2\gamma^2}{\eta} + \frac{4\gamma^2 - \gamma^4}{\eta^3}, \nonumber\\
\Omega^-_1 & \approx &  \frac{\gamma^2}{2\eta^2} + \frac{2\gamma^2 - \gamma^4}{2\eta^4},     \nonumber\\
\Omega^-_2 & \approx & -\frac{\gamma}{2\eta} + \frac{\gamma^3 - 4\gamma}{4\eta^3},           \nonumber\\
\Omega^-_3 & \approx &  \frac{\gamma}{\eta^2}+\frac{4\gamma - 3\gamma^3}{2\eta^4}.
\end{eqnarray}
It follows that
\begin{equation}\label{largeBC41}
{\cal C}^{(4)}[|\Phi^{15}\rangle\langle\Phi^{15}|] \approx \frac{2\gamma^2}{\eta^2} + \frac{8\gamma^2 - 4\gamma^4}{\eta^4},
\end{equation}
\begin{equation}\label{largeBC42}
{\cal C}_4[|\Phi^{15}\rangle\langle\Phi^{15}|] \approx \frac{2\gamma}{\eta} + \frac{24\gamma - 9\gamma^3}{4\eta^3}.
\end{equation}
We note that $|\Phi^{15}\rangle$ is a state with genuine four-partite entanglement \cite{comment1}, which remains non-zero even for large $\eta$ and going to zero only in the asymptotic limit of infinite magnetic field.

Secondly, we have
\begin{eqnarray}
H_4|\Phi^{14}\rangle & = & -[(\alpha^+ + \alpha^-)\gamma + 2]J|\Phi^{14}\rangle, \nonumber \\
H_4|\Phi^{13}\rangle & = & -[(\alpha^+ + \alpha^-)\gamma - 2]J|\Phi^{13}\rangle,
\end{eqnarray}
where
\begin{equation}
\alpha^{\pm} \equiv \frac{\sqrt{\eta^2+4\gamma^2} \pm \eta}{2\gamma},
\end{equation}
and
\begin{eqnarray}
|\Phi^{14}\rangle 
& = & \frac{1}{2\sqrt{1 + (\alpha^{-})^2}}(-\alpha^-\ket{0001} + \alpha^-\ket{0010} - \alpha^-\ket{0100} + \ket{0111} \nonumber \\
& & \,\,\,\,\,\,\,\,\,\,\,\,\,\,\,\,\,\,\,\,\,\,\,\,
+ \alpha^-\ket{1000} - \ket{1011} + \ket{1101} - \ket{1110}), \nonumber \\
|\Phi^{13}\rangle 
& = & \frac{1}{2\sqrt{1 + (\alpha^{-})^2}}(-\alpha^-\ket{0001} - \alpha^-\ket{0010} - \alpha^-\ket{0100} + \ket{0111} \nonumber \\
& & \,\,\,\,\,\,\,\,\,\,\,\,\,\,\,\,\,\,\,\,\,\,\,\,
- \alpha^-\ket{1000} + \ket{1011} + \ket{1101} + \ket{1110}).
\end{eqnarray}
Corresponding to these states are $|\Phi^1\rangle$ and $|\Phi^2\rangle$, which can be derived from $|\Phi^{14}\rangle$ and $|\Phi^{13}\rangle$ respectively by $-\alpha^- \rightarrow \alpha^+$.  They satisfy $H_4|\Phi^{1, 2}\rangle = [(\alpha^+ + \alpha^-)\gamma \pm 2]J|\Phi^{1, 2}\rangle$.  The MKB concurrences for $|\Phi^{14}\rangle$ are given by
\begin{eqnarray}
{\cal C}^{(4)}[\ketbra{\Phi^{14}}{\Phi^{14}}] & = & \frac{2\alpha^-}{1+(\alpha^{-})^2}, \nonumber \\
{\cal C}_4[\ketbra{\Phi^{14}}{\Phi^{14}}] & = & \frac{1}{1 + (\alpha^-)^2}\sqrt{\frac{3 + 8(\alpha^-)^2 + 3(\alpha^-)^4}{2}}.
\end{eqnarray}
Hence, $|\Phi^{14}\rangle$ is also a state with genuine four-partite entanglement \cite{osterloh05, Osterloh05}.  It reduces to a W state when $\gamma = 0$.

Thirdly, we have $H_4|\Phi^{3,4}\rangle = \eta J|\Phi^{3,4}\rangle$ and $H_4|\Phi^{11,12}\rangle = -\eta J|\Phi^{11,12}\rangle$, where
\begin{eqnarray}
\ket{\Phi^{3, 4}}   & = & \frac{1}{2}(\ket{0001} \pm \ket{0010} - \ket{0100} \mp \ket{1000}), \nonumber \\
\ket{\Phi^{11, 12}} & = & \frac{1}{2}(\ket{0111} \pm \ket{1011} - \ket{1101} \mp \ket{1110}).
\end{eqnarray}
These states belong to the family of W states which are entangled but do not contain genuine four-partite entanglement \cite{osterloh05}.  Fourthly, by substituting $\omega^+$ in Eq.(\ref{Omega}) with $\omega^-$ we obtain the corresponding $\Delta^{\pm}_1$, $\Delta^{\pm}_2$ and $\Delta^{\pm}_3$ in terms of which we express
\begin{eqnarray}
\ket{\Phi^{10,11}}& = & 
N_\Delta^\pm(\Delta_1^\pm\ket{0000}+\Delta_2^\pm\ket{0011}+\Delta_3^\pm\ket{0101}+\Delta_2^\pm\ket{0110}\nonumber\\
& & + \Delta_2^\pm\ket{1001}+\Delta_3^\pm\ket{1010}+\Delta_2^\pm\ket{1100}+\ket{1111}).
\end{eqnarray}
Here, $N_\Delta^\pm \equiv 1/\sqrt{1+(\Delta_1^\pm)^2+4(\Delta_2^\pm)^2+2(\Delta_3^\pm)^2}$.  They satisfy $H_4|\Phi^{10,11}\rangle = \pm\omega^-J|\Phi^{10, 11}\rangle$.  Lastly, we have the following four degenerate eigenstates with eigenvalue zero:
\begin{eqnarray}
\ket{\Phi^6} & = & \frac{1}{\sqrt{2}}\left(\ket{0011}-\ket{1100}\right), \nonumber \\
\ket{\Phi^7} & = & \frac{1}{\sqrt{2}}\left(\ket{0101}-\ket{1010}\right), \nonumber \\
\ket{\Phi^8} & = & \frac{1}{\sqrt{2}}\left(\ket{0110}-\ket{1001}\right), \nonumber \\
\ket{\Phi^9} & = & \frac{1}{2}\left(\ket{0011}-\ket{0110}-\ket{1001}+\ket{1100}\right).
\end{eqnarray}
The eigenstates $\ket{\Phi^{6,7,8}}$ belong to the family of GHZ states that contain only genuine four-partite entanglement, while $\ket{\Phi^9}$ is a product state of two Bell states.

As in the case of the two-qubit model, we construct the thermal state of the four-qubit Heisenberg $XY$ chain as follows:
\begin{equation}\label{thermalstate4}
\chi_4 = \sum^{15}_{i = 0}w_i|\Phi^i\rangle\langle\Phi^i| = \frac{1}{Z_4}\mathrm{e}^{-\beta H_4}.
\end{equation}
Here, the partition fuction
\begin{eqnarray}
Z_4 & = & 4 + 4\cosh\beta\eta J + 2\cosh\beta[(\alpha^+ + \alpha^-)\gamma + 2]J \nonumber \\
& & + 2\cosh\beta[(\alpha^+ + \alpha^-)\gamma - 2]J + 2\cosh\beta\omega^+J + 2\cosh\beta\omega^-J.
\end{eqnarray}
At non-zero temperatures the state of the system becomes a mixture of the energy eigenstates with statistical weights
\begin{equation}
w_{15} = \frac{e^{\beta\omega^+J}}{Z_4},\ w_{14} = \frac{e^{\beta[(\alpha^+ + \alpha^-)\gamma + 2]J}}{Z_4},\ \cdots
\end{equation}

\subsection{Zero Temperature}
\begin{figure}[h!]
\begin{center}
\includegraphics[scale=0.25]{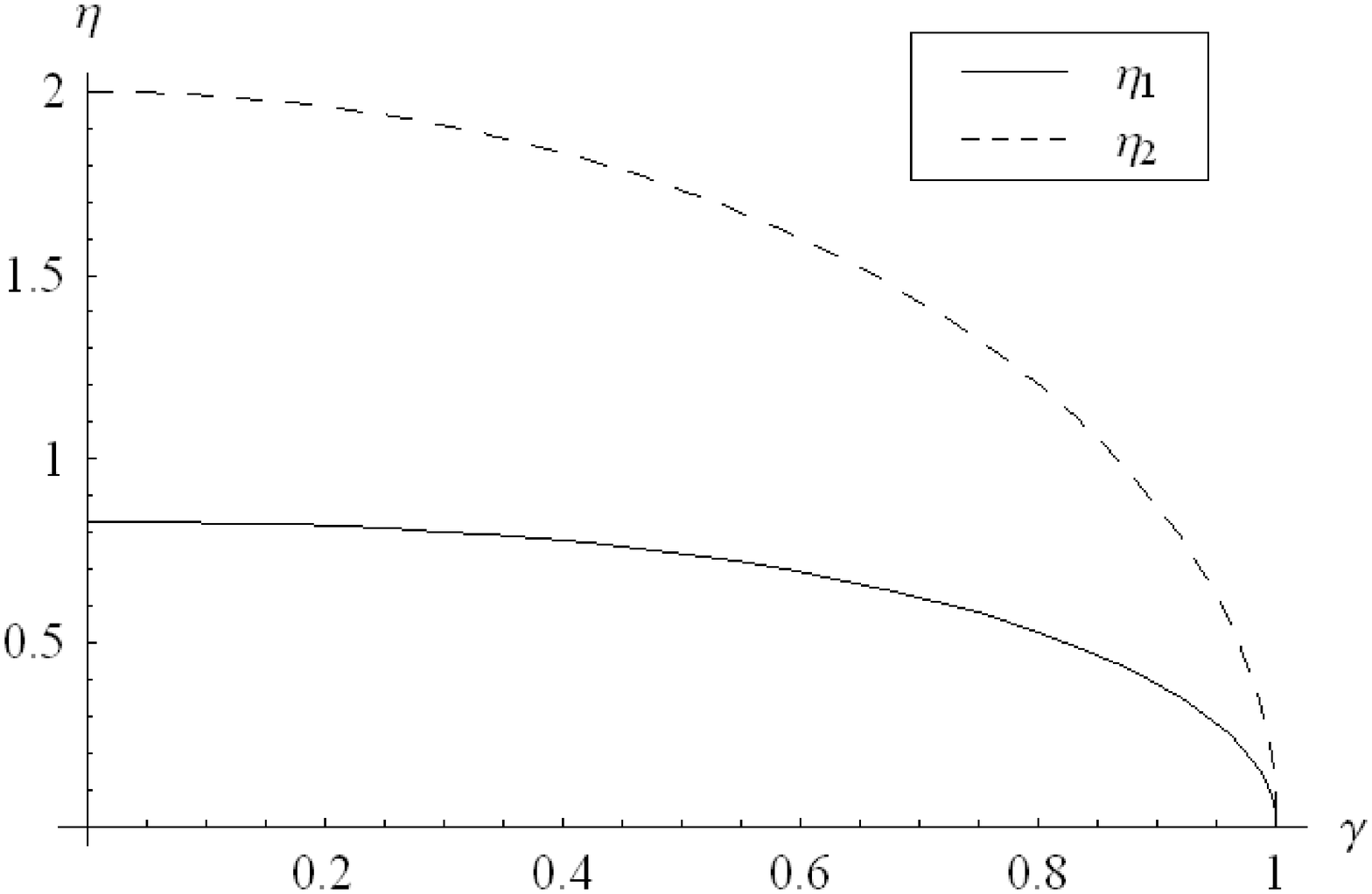}
\caption{\label{transmag}
The two transition $\eta$'s for the system at zero temperature plotted against the anisotropy $\gamma$.}
\end{center}
\end{figure}

In the $\beta \rightarrow \infty$ limit, 
\begin{equation}
\chi_4 = \frac{1}{Z_4}
[e^{\beta[(\alpha^+ + \alpha^-)\gamma + 2]J}\ketbra{\Phi^{14}}{\Phi^{14}} + e^{\beta\omega^+J}\ketbra{\Phi^{15}}{\Phi^{15}}],
\end{equation}
with $Z_4 = e^{\beta[(\alpha^+ + \alpha^-)\gamma + 2]J} + e^{\beta\omega^+J}$.  Like in the 2-qubit Heisenberg XY model, there are thus two possible lowest energy states, namely $\ket{\Phi^{14}}$ or $\ket{\Phi^{15}}$, depending on the strength of the applied magnetic field.  As $\eta$ is increased from zero, there are in general two instances when $(\alpha^+ + \alpha^-)\gamma + 2 = \omega^+$.  We let $\eta_1$ and $\eta_2$ denote the solutions.  Their dependence on the anisotropy of the system are plotted in Fig. \ref{transmag}.  As $\gamma$ is increased from zero, both $\eta_1$ and $\eta_2$ become smaller and converge to zero when $\gamma = 1$.  In fact, for the Ising model ($\gamma = 1$), both $\ket{\Phi^{14}}$ and $\ket{\Phi^{15}}$ are the lowest energy states of the system.  The system state is an equal mixture of both states:
\begin{equation}\label{equalmix}
\chi_4 = \frac{1}{2}[\ketbra{\Phi^{14}}{\Phi^{14}} + \ketbra{\Phi^{15}}{\Phi^{15}}].
\end{equation}
But, once the external magnetic field is turned on, $\ket{\Phi^{15}}$ becomes the only ground state of the model regardless of the strength of the field.

\begin{figure}[h!]
\begin{center}
\includegraphics[scale=0.3]{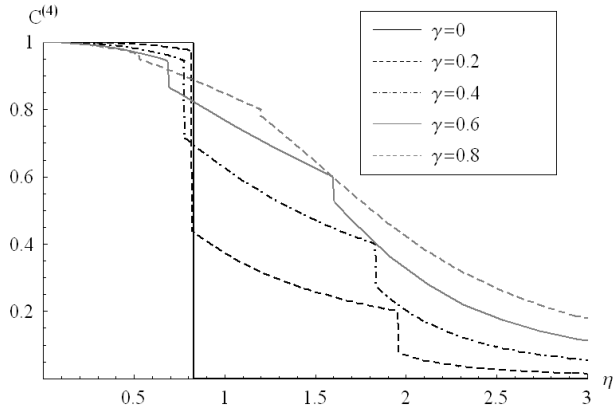}
\caption{\label{genuine}
The MKB concurrence ${\cal C}^{(4)}$, which is an entanglement monotone, plotted against magnetic field $\eta$ for different values of anisotropy $\gamma$.  In general, there are sharp changes in ${\cal C}^{(4)}$ at the two transition fields $\eta_1$ and $\eta_2$.}
\end{center}
\end{figure}

\begin{figure}[h!]
\begin{center}
\includegraphics[scale=0.3]{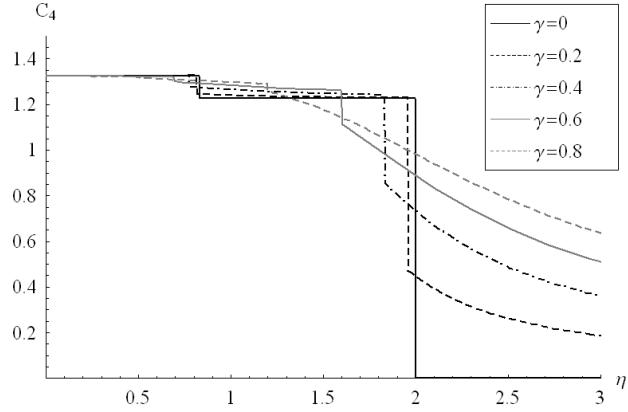}
\caption{\label{confour}
The MKB concurrence ${\cal C}_4$ plotted against $\eta$ for different values of $\gamma$.  Similar to Fig.\ref{genuine}, there are sharp changes of the concurrence at the two transition magnetic fields.}
\end{center}
\end{figure}

It follows that for $0 < \gamma < 1$, depending on $\eta$ we have the density operator of the system
\begin{equation}\label{gs}
\chi_4 = \left \{\begin{array}{ll}
\ketbra{\Phi^{15}}{\Phi^{15}} & 0 \leq \eta < \eta_1 \\
\frac{1}{2}[\ketbra{\Phi^{15}}{\Phi^{15}} + \ketbra{\Phi^{14}}{\Phi^{14}}] & \eta = \eta_1 \\
\ketbra{\Phi^{14}}{\Phi^{14}} & \eta_1 < \eta < \eta_2 \\
\frac{1}{2}[\ketbra{\Phi^{14}}{\Phi^{14}} + \ketbra{\Phi^{15}}{\Phi^{15}}] & \eta = \eta_2 \\
\ketbra{\Phi^{15}}{\Phi^{15}} & \eta > \eta_2
\end{array}\right.
\end{equation}
We can therefore calculate the MKB concurrrences of the system at zero temperature ($\beta\rightarrow\infty$) for the different regions of magnetic field strength. Plots of ${\cal C}^{(4)}$ and ${\cal C}_4$ against magnetic field for different values of anisotropy $\gamma$ are shown in Fig. \ref{genuine} and \ref{confour}.  For both concurrences, there are sharp changes at the transition magnetic fields due to quantum phase transition.

\subsection{Non-zero Temperatures}
\begin{figure}[h!]
\begin{center}
\includegraphics[scale=0.5]{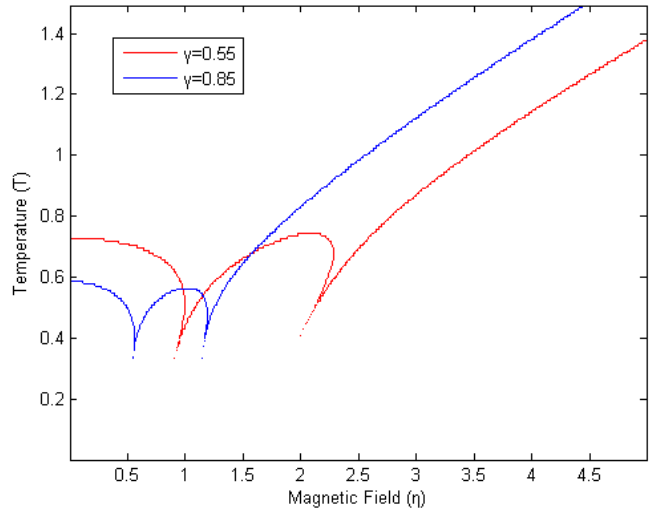}
\caption{\label{region4}
The points where the MKB concurrence ${\cal C}^{(4)}$ equals zero are plotted for two different values of $\gamma$.  The region above each curve is the region where ${\cal C}^{(4)}=0$.}
\end{center}
\end{figure}

In general, as the temperature of a system is increased, its density operator becomes closer to the maximally mixed state, $\frac{1}{n}I$, where $n$ is the dimension of the Hilbert space and $I$ is the identity operator.  There thus exist critical temperatures $T_c$ beyond which the MKB concurrences of the system become zero, like in the two-qubit case (see Fig. \ref{region4}).  The existence of $T_c$'s is guaranteed by the fact that a state becomes separable when it is sufficiently close to $\frac{1}{n}I$ \cite{zyczkowski}.

On the other hand, it can also be seen from Fig. \ref{region4} that it is always possible to have non-zero entanglement in the system by applying a sufficiently large magnetic field to it.  In particular, even when $T > T_c$, one could reintroduce entanglement into the system by increasing $\eta$.  This is always possible for finite temperatures as long as $\gamma\neq0$.  Hence, the thermal entanglement associated with the four-qubit Heisenberg $XY$ model exhibits the same characteristic robustness as in the two-qubit case.  We note that, in contrast to the two-qubit case (see Fig. 3 in Ref.\cite{kamta02}), each of our graphs in Fig. \ref{region4} has two ``singular turning points''.  This is due to the fact that there are two transition $\eta$'s instead of one in the two-qubit model.

In order to understand this robustness, we draw inspiration from the two-qubit case.  We observe that for any finite temperature $T$, the statistical weight $w_{15}$ when $\eta$ is large enough, is given by 
\begin{equation}\label{lw}
w_{15} = \frac{1}{Z_4}\exp\beta\omega^+J = \frac{1}{\xi}
\end{equation}
where $\xi \equiv 1 + \mathrm{e}^{-2\beta\omega^+J} + 4\mathrm{e}^{-\beta\omega^+J} + 4\mathrm{e}^{-\beta\omega^+J}\cosh\beta\eta J + 2\mathrm{e}^{-\beta\omega^+J}\cosh\beta[(\alpha^+ + \alpha^-)\gamma + 2]J + 2\mathrm{e}^{-\beta\omega^+J}\cosh\beta[(\alpha^+ + \alpha^-)\gamma - 2]J + 2\mathrm{e}^{-\beta\omega^+J}\cosh\beta\omega^-J$.  It can be shown that the large $\eta$ behavior of $w_{15}$ is independent of $\gamma$.  Furthermore, by increasing $\eta$ appropriately, $\xi$ can be made to be as close as possible to unity.  This results in $\ket{\Phi^{15}}$ being the only state that significantly contributes to the thermal state $\chi_4$.  In the light of Eq.(\ref{approx}), we may conclude that
\begin{equation}
{\cal C}^{(4)}[\chi_4] \approx {\cal C}^{(4)}[|\Phi^{15}\rangle\langle\Phi^{15}|].
\end{equation}
That this is indeed the case has been established numerically (see Table \ref{comc4}).  Since the MKB concurrence ${\mathcal C}^{(4)}$ can be determined exactly, it is calculated for states $\chi_4$ and $\ket{\Phi^{15}}$ under different combinations of $\eta$ and $T$ (some results are shown in Table \ref{comc4}).  The two values for each combination of $\eta$ and $T$ are then compared.  For a given temperature, the two values, ${\cal C}^{(4)}[\chi_4]$ and ${\cal C}^{(4)}[\ketbra{\Phi^{15}}{\Phi^{15}}]$, agree when large enough magnetic field is applied.

\begin{table}[ht!]
\begin{center}
\begin{tabular}{|c|c|c|c|c|c|c|}\hline
&\multicolumn{6}{|c|}{${\cal C}^{(4)}$}\\\cline{2-7}
$T$&\multicolumn{2}{|c|}{$\eta=$0}&\multicolumn{2}{|c|}{$\eta=$100}&\multicolumn{2}{|c|}{$\eta=$1000}\\\cline{2-7}
&$\chi_4$&$|\Phi^{15}\rangle\langle\Phi^{15}|$&$\chi_4$&$|\Phi^{15}\rangle\langle\Phi^{15}|$
&$\chi_4$&$|\Phi^{15}\rangle\langle\Phi^{15}|$\\\hline
1&0&1&0.0000180069&0.0000180069&1.79177$\times$10$^{-7}$&1.79177$\times$10$^{-7}$\\\hline
5&0&1&0.0000180068&0.0000180069&1.79177$\times$10$^{-7}$&1.79177$\times$10$^{-7}$\\\hline
10&0&1&0.0000174316&0.0000180069&1.79177$\times$10$^{-7}$&1.79177$\times$10$^{-7}$\\\hline
50&0&1&0&0.0000180069&1.79175$\times$10$^{-7}$&1.79177$\times$10$^{-7}$\\\hline
100&0&1&0&0.0000180069&1.07513$\times$10$^{-7}$&1.79177$\times$10$^{-7}$\\\hline
\end{tabular}
\caption{A comparison between ${\cal C}^{(4)}[\chi_4]$ and ${\cal C}^{(4)}[\ketbra{\Phi^{15}}{\Phi^{15}}]$ for some combinations of $\eta$ and $T$.}\label{comc4}
\end{center}
\end{table}

Hence, for $\gamma\neq0$, the revival and robustness of entanglement in the Heiseberg four-qubit model can be understood, as in the two-qubit case, in terms of the large $\eta$ behaviors of both the MKB concurrence ${\mathcal C}^{(4)}[\ketbra{\Phi^{15}}{\Phi^{15}}]$ (Eq.(\ref{largeeta})) and the statistical weight $w_{15}$ (Eq.(\ref{lw})).  We may conclude, when $w_{15} \approx 1$, that the entanglement associated with $\chi_4$ is of the genuine four-partite kind \cite{osterloh05, Osterloh05}.  The total entanglement as measured by ${\mathcal C}_4$ thus also undergoes revival since genuine four-partite entanglement is only one kind of the entanglement measured by ${\mathcal C}_4$.  In contrast, for $\gamma=0$ and large $\eta$, the ground state is the product state $\ket{1111}$.  This is an important distinction between the four-qubit {\it XX} model and the {\it XY} model.  Consequently, no revival of entanglement is observed for the case of four-qubit isotropic Heisenberg $XX$ chain.

We may apply the above analysis employing ${\cal C}^{(n)}$ to study Heisenberg $XY$ chains with even number $n$ of qubits.  Firstly, we determine if the ground state $|\Phi^g\rangle$ remains genuinely multipartite entangled at large $\eta$.  Secondly, we determine if the ground state statistical weight $w_g$ can be made very close to 1 at large $\eta$.  Through Eq.(\ref{approx}), these two properties together imply that for any finite temperature, nonzero genuine multipartite entanglement can always be obtained by applying a large enough magnetic field.  We show that the six-qubit Heisenberg $XY$ chain has both properties in the next section.


\section{Six-qubit Heisenberg $XY$ model and beyond}
In this section, we show that the six-qubit Heisenberg $XY$ chain (with $\gamma\neq0$) does indeed possess the characteristic properties, which enable the model to have non-zero thermal entanglement at any given finite temperature when subject to an external magnetic field of appropriate strength.  To this end, we determine the eigenstate $|\Phi^g\rangle$ of $H_6$ whose eigenvalue $E_g = -\lambda J$ is the minimum when $\eta$ is large.  Here,
\begin{equation}
\lambda \equiv \sqrt{3(2+2\gamma^2+\eta^2)+2\kappa+2\sqrt{2}\sqrt{(4\gamma^2+\eta^2)((3+\gamma^2+\eta^2)+\kappa)}}
\end{equation}
with $\kappa \equiv \sqrt{\gamma^4+(-3+\eta^2)^2+2\gamma^2(3+\eta^2)}$.  And,
\begin{eqnarray}
\ket{\Phi^g}& = & N
(\ket{000000}\Theta_1+\ket{000011}\Theta_2+\ket{000101}\Theta_3+\ket{000110}\Theta_4 + \ket{001001}\Theta_4+\ket{001010}\Theta_3
\nonumber\\ & & 
+\ket{001100}\Theta_2+\ket{001111}\Theta_5+\ket{010001}\Theta_3+\ket{010010}\Theta_4 + \ket{010100}\Theta_3+\ket{010111}\Theta_6
\nonumber\\ & & 
+\ket{011000}\Theta_2+\ket{011011}\Theta_7+\ket{011101}\Theta_6+\ket{011110}\Theta_5+\ket{100001}\Theta_2+\ket{100010}\Theta_3
\nonumber\\ & & 
+\ket{100100}\Theta_4+\ket{100111}\Theta_5+\ket{101000}\Theta_3+\ket{101011}\Theta_6+\ket{101101}\Theta_7+\ket{101110}\Theta_6
\nonumber\\ & & 
+\ket{110000}\Theta_2+\ket{110011}\Theta_5+\ket{110101}\Theta_6+\ket{110110}\Theta_7+\ket{111001}\Theta_5+\ket{111010}\Theta_6
\nonumber\\ & & 
+\ket{111100}\Theta_5+\ket{111111}\Theta_8)
\end{eqnarray}
with
\begin{eqnarray}
\Theta_1&=&\frac{\{24J^2-(\lambda-3J\eta)(\lambda-J\eta)\}\Theta_8+2J^2(\lambda-J\eta)\zeta+8J^2\tau}{2J\gamma(\lambda+3J\eta)},\nonumber\\
\Theta_2&=&-\frac{(\lambda+3J\eta)\Theta_1}{6J\gamma},\nonumber\\
\Theta_3&=&-\frac{(\lambda+3J\eta)\Theta_8+(\lambda-J\eta)\tau+2J^2\zeta}{6J\gamma^2},\nonumber\\
\Theta_4&=&\frac{3(2-\gamma^2)\Theta_8+(\lambda-J\eta)\zeta+2\tau}{3\gamma^2},\nonumber\\
\Theta_5&=&-\frac{(\lambda-3J\eta)\Theta_8}{6J\gamma},\nonumber\\
\Theta_6&=&\frac{3\Theta_8+\tau}{3\gamma},\nonumber\\
\Theta_7&=&-\frac{(\lambda-3J\eta)\Theta_8-2J^2\zeta}{6J\gamma},\nonumber\\
\Theta_8&=&\lambda^4-2J^2\lambda^2(2+2\gamma^2+\eta^2)+J^4\{\eta^2(\eta^2-12)+4\gamma^2(\eta^2+4)\},\nonumber\\
N&=&\frac{1}{\sqrt{\Theta_1^2+\Theta_8^2+3(2\Theta^2_2+2\Theta^2_3+\Theta^2_4+2\Theta^2_5+2\Theta^2_6+\Theta^2_7)}},
\end{eqnarray}
and
\begin{eqnarray}
\tau&=&-8J\lambda\eta\{\lambda^2-J^2(6-2\gamma^2+\eta^2)\},\\
\zeta&=&4\lambda\{\lambda^2(\gamma^2-1)-J^2(4\gamma^4-9\eta^2-4\gamma^2+\gamma^2\eta^2)\}.
\end{eqnarray}
The MKB concurrence ${\cal C}^{(6)}$ is calculated for this state, giving
\begin{equation}
{\cal C}^{(6)}[\ketbra{\Phi^g}{\Phi^g}] = 2N^2|\Theta_1\Theta_8 + 6\Theta_2\Theta_5 + 6\Theta_3\Theta_6 + 3\Theta_4\Theta_7|.
\end{equation}
In the limit of large $\eta$,
\begin{equation}
{\cal C}^{(6)}[\ketbra{\Phi^g}{\Phi^g}] \approx 2\gamma^3\eta^{-3}.
\end{equation}
It can again be shown that $|\Phi^g\rangle$ is a genuine six-partite entangled state \cite{osterloh05, Osterloh05}.  In addition, the statistical weight $w_g$ of the ground state $|\Phi^g\rangle$ can be shown numerically to be close to unity when an appropriately large magnetic field is applied.  The ground state $|\Phi^g\rangle$ therefore possesses the two desired properties for the system to have non-zero genuine six-partite entanglement at any finite temperature, provided an appropriate magnetic field is applied.  We conjecture that the robustness of genuine multipartite entanglement is a general property of Heisenberg $XY$ chain with even number of particles.  This conjecture can be experimentally tested, as will be discussed in the next section. 


\section{Conclusions}
In this study, we have investigated in detail the origin of the robustness of genuine multipartite entanglement in two-, four-, and six-qubit Heisenberg $XY$ models.  Two important properties possessed by the ground states $|\Phi^g\rangle$ of these models, which enable them to exhibit the characteristic robustness, were identified.  The first property, namely the statistical weight $w_g$ associated with $|\Phi^g\rangle$ can be made very close to unity by applying a large enough magnetic field, allows us to use Eq.(\ref{approx}) and conclude that the MKB concurrence of the thermal state $\chi$ equals that of the ground state.  It follows that we could obtain fairly accurately the MKB concurrence of $\chi$ (a mixed state) by calculating the MKB concurrence of $|\Phi^g\rangle$ (a pure state) in this case.  The second property being that $|\Phi^g\rangle$ remains genuinely multipartite entangled under such a magentic field, then guarantees that there is non-zero genuine multipartite entanglement for any finite temperature as long as sufficiently large magnetic field is applied.  These properties allow us to extend our study to $XY$ chains with any even number of qubits.  Heisenberg $XY$ chains with odd number of qubits were not considered in this study because ${\cal C}^{(n)} = 0$ for odd $n$.  However, one could similarly study the robustness of other kinds of entanglement employing the other MKB concurrences.

In fact, the applicability of our analysis is not only restricted to the Heisenberg $XY$ chains.  As long as the ground state of a system can be made to dominate the thermal state at any finite temperature by adjusting some parameters of the system Hamiltonian, Eq.(\ref{approx}) tells us that the MKB concurrence of the thermal state is given by that of the ground state.  So, if in addition, the ground state remains multipartite entangled under these conditions, similar robustness of multipartite entanglement is expected to be observed.  Therefore, our analysis can be used to identify possible candidates for realization of quantum computation at finite temperatures.

The MKB concurrences ${\mathcal C}^{(n)}$ and ${\mathcal C}_n$ can be directly measured for pure states, if two copies of the states are available \cite{Walborn, aolita06}.  This is obviously due to the fact that the  MKB concurrences are defined in terms of expectation values of Hermitean operators.  Indeed, this has lead to the direct measurement of Wootters concurrence, previously thought to be not directly measurable, for pure states in laboratory \cite{Walborn}.  Since we are interested in the region where the ground state $|\Phi^g\rangle$ is the only state that contributes significantly to the thermal state $\chi$, the state $\chi$ is ``almost pure" and therefore it is possible to measure the MKB concurrences of $\chi$ with a high degree of success.  The results obtained here can thus be experimentally verified.

In conclusion, we have established a rather general method of identifying systems that exhibit robustness of multipartite entanglement at finite temperatures.  Our analysis rest on the mathematical property of the MKB concurrences, namely Eq.(\ref{approx}).  The MKB concurrences can be directly measured and therefore experiments can be carried out to verify the results of any analysis.



\begin{thebibliography}{99}
\bibitem{comment1} It turns out that ${\cal C}^{(4)}[|\Psi\rangle\langle\Psi|] \not= 0$ is only a sufficient condition for $|\Psi\rangle$ to have genuine multipartite entanglement.  However, in this case, it can be shown that $|\Phi^{15}\rangle$ possesses genuine four-partite entanglement \cite{osterloh05, Osterloh05}.
\end{thebibliography}
\end{document}